\documentclass[5p,final]{elsarticle}
\usepackage{subfigure}
\usepackage{amsmath}
\usepackage{siunitx}
\usepackage{blindtext}
\usepackage{graphicx}
\usepackage{verbatim}
\biboptions{longnamesfirst,semicolon}
\title{Effect of antimony substitution in iron pnictide compounds}

\author[bt]{D. Schmidt}
\ead{daniel.schmidt@uni-bayreuth.de}
\author[bt]{H. F. Braun}
\address[bt]{Physikalisches Institut, Lehrstuhl Experimentalphysik V,
Universit\"at Bayreuth, 95440 Bayreuth}

\newcommand{\chem}[1]{$_{\mathrm{#1}}$}
\newcommand{\chemex}[2]{$_{\mathrm{#1}-\mathrm{#2}}$}

\begin{document}

\begin{abstract}
In the present study we have examined the effect of negative chemical pressure in iron pnictides. We have synthesized substitution series replacing arsenic by antimony in a number of 1111- and 122-iron arsenides and present their crystallographic and physical properties. The SDW transition temperature in LaFeAs\chemex{1}{x}Sb\chem{x}O decreases with increasing antimony content, while
the superconducting transition temperature in LaFeAs\chemex{1}{x}Sb\chem{x}O\chem{0.85}F\chem{0.15} initially increases with Sb substitution. 1111-compounds with samarium instead of lanthanum have a smaller unit cell volume. In these phases, no Sb solubility is observed. There is also no apparent solubility of antimony in the 122-iron arsenides.
\end{abstract}

\maketitle

\section{Introduction}

The discovery of superconductivity in fluorine doped LaFeAsO (1111) compounds \cite{kamihara_2008} was followed by quite a large amount of experimental and theoretical work, which was aimed at understanding the physical properties of iron pnictides. Beside the superconducting 1111 iron pnictides which crystallize in the tetragonal ZrCuSiAs-type structure, superconductivity was found in the 122 compounds which crystallize in the ThCr\chem{2}Si\chem{2} structure type \cite{rotter_superconductivity_2008} and the 111 compounds with PbFCl-type structure \cite{tapp_lifeas:_2008}. In the present study we concentrate on the 1111-type compounds with formula REFeAsO (RE: rare earth) and 122-type compounds with formula AEFe$_{2}$As$_{2}$ (AE: alcaline earth). Both structure types are characterized by layers of FeAs which are separated by layers of rare earth oxide or of alcaline earth atoms.

Upon cooling, the undoped compounds undergo a structural phase transition from tetragonal to orthorhombic. Beside this structural transition, there is also a magnetic phase transition to long range antiferromagnetic order. This transition is ascribed to the occurrence of a spin-density-wave (SDW) \cite{rotter_spin-density-wave_2008}. In the 122-compounds, the structural and magnetic phase transition takes place at the same temperature whereas in the 1111-compounds the transition temperature of the structural transition is somewhat higher than the magnetic transition temperature \citep{klauss_commensurate_2008}. 

By appropriate doping, it is possible to suppress the SDW and as a consequence superconductivity (SC) may arise. There are several concepts for doping. In the 122-compounds, hole doping on the alcaline earth site by substitution with potassium or sodium results in the highest superconducting transition temperature $T_\mathrm{c} = 38$\,K for Ba$_{0.6}$K$_{0.4}$Fe$_{2}$As$_{2}$ \cite{rotter_superconductivity_2008, aswartham_hole_2012}. In the 1111-compounds, hole doping can be achieved by substituting the rare earth by an alcaline earth. For example, the substitution with Sr induces superconductivity with a maximum $T_\mathrm{c}$ of \SI{25}{\kelvin} \citep{wen_superconductivity_2008}. In both types of compounds, electron doping was achieved by replacing Fe with e.g. Co  \cite{sefat_superconductivity_2008}, Ni \cite{sefat_structure_2009}, Ru \cite{sharma_superconductivity_2010}, Pd or Rh \cite{ni_phase_2009} which leads to superconductivity with a maximum $T_\mathrm{c}$ = 25\,K for BaFe$_{2-x}$Co$_{x}$As$_{2}$ \cite{wang_peculiar_2009}. By substituting the iron with cobalt in the 1111-compounds, superconductivity was obtained with transition temperatures up to 17\,K \citep{sefat_superconductivity_2008_1111, wang_effects_2009}. These atomic substitutions of an alkaline element for AE, AE for RE or iron by some other transition element of neighbor groups not only changes the electron concentration but at the same time leads to a change in unit cell volume. The effect of such a volume change alone can be observed by the application of hydrostatic pressure.

In both compounds, BaFe$_{2}$As$_{2}$ and LaFeAsO, superconductivity can be induced under high pressure, where BaFe$_{2}$As$_{2}$ reaches a maximum $T_\mathrm{c}$ of 29\,K at a pressure of \SI{4}{\giga \pascal} \cite{alireza_superconductivity_2009} while for LaFeAsO a maximum $T_\mathrm{c}$ of \SI{21}{\kelvin} is reached at \SI{12}{\giga \pascal} \citep{okada_superconductivity_2008}. 

Similar effects might be obtained by ``chemical pressure'', i.e. a change in unit cell volume by isoelectronic substitution on one of the lattice sites. Positive chemical pressure results from the substitution of As through the smaller P. In such substitution the SDW can be suppressed and superconductivity occurs with a maximum $T_\mathrm{c}$ = \SI{31}{\kelvin} for BaFe$_{2}$As$_{1.74}$P$_{0.26}$ \cite{kasahara_evolution_2010} and up  to $T_\mathrm{c}$ = \SI{11}{\kelvin} for LaFeAs$_{1-x}$P$_{x}$O ($x = 0.25-0.3$) \citep{wang_superconductivity_2009}. 

The possible substitution of Sb for As, corresponding to negative chemical pressure, has been examined in theoretical studies. First principle calculations were performed by density-functional methods on  compounds which had not been synthesized then. In the 1111-compounds the substitution was predicted to cause an enhanced Fermi surface nesting and the 1111-compound with antimony is described as a candidate for higher $T_\mathrm{c}$ values \cite{moon_enhanced_2008}. In the 122-compounds no enhanced nesting is predicted \cite{moon_dominant_2009}. For both the 1111- and 122-compounds with antimony, expected lattice parameters have been calculated \cite{moon_enhanced_2008, moon_dominant_2009, lebegue_delicate_2009}. 

Experimentally, it was possible to substitute the As by Sb in the LaFeOAs compounds. With increasing Sb content, there is a decrease of the SDW and structural phase transition temperature, however, no superconductivity is induced \cite{carlsson_effect_2011}.  To complement the investigations on the fluorine free 1111-compounds we have synthesized both the lanthanum and samarium compounds. 

Wang et al. reported on Sb substitution in fluorine doped 1111-samples 
and found a recovery of the SDW transition in LaFeAs\chemex{1}{x}Sb\chem{x}O\chem{0.9}F\chem{0.1} with $x > 0.1$ \cite{wang_structural_2010}. In order to expand the phase diagram of the fluorine doped system, we prepare an antimony substitution series with a fluorine content of 0.15. So far, to the best of our knowledge, there are no experimental results available about the possibility of Sb substitution in BaFe$_{2}$As$_{2}$. In order to compare the 1111- and 122-compounds with respect to the influence of Sb substitution on the SDW transition as well as on the superconducting transition we used non-superconducting BaFe$_{2}$As$_{2}$ and superconducting BaFe$_{1.83}$Co$_{0.17}$As$_{2}$ as parent compounds for our Sb substitution experiments with the 122-compounds. 

\section{Materials and Methods}

Due to the air sensitivity of the Ba-containing precursor, all preparation steps were carried out in a glove box filled with dry Ar-gas. The synthesis of the polycrystalline samples was carried out in sealed quartz glass tubes under inert atmosphere, by solid state reaction of the precursors BaAs, Fe$_2$As (Alfa Aesar \SI{99.5}{\percent}), Co$_2$As and Fe$_2$Sb, respectively. The precursor BaAs was synthesized by a vapour transport method. It was important that Ba and As were not in direct contact since their highly exothermic reaction lead to a rapid increase of vapour pressure, even at moderate temperature, which could  break the quartz tube. 

To avoid such uncontrolled reactions, Ba chunks (Alfa Aesar \SI{99.2}{\percent}) were kept in a corundum crucible inside a quartz glass container while As pellets (Alfa Aesar \SI{99.99}{\percent}) in a 1:1 molar ratio were placed underneath or around this crucible. The tube was evacuated, sealed under Ar at a pressure of \SI{0.05}{\bar} and heated up slowly in a box-type furnace to \SI{973}{\kelvin}. This temperature was kept for \SI{5}{\day}. After the heat treatment, the As had evaporated and quantitatively reacted with the Ba. The reaction results in dark grey lumps with mean composition BaAs. The Co$_{\mathrm{2}}$As and Fe$_{\mathrm{2}}$As precursors were prepared by slowly heating  stoichiometric mixtures of the elements (Fe powder Gr\"ussing \SI{99.9}{\percent}, Co powder Alfa Aesar \SI{99.8}{\percent}, Sb shot Alfa Aesar \SI{99.9999}{\percent}) and keeping them at \SI{973}{\kelvin} for \SI{48}{\hour} in an evacuated quartz capsule. In order to synthesize the substitution series, the precursors were ground, mixed together in the appropriate ratio, pressed into pellets and heated at a rate of \SI{100}{\kelvin \per \hour} up to \SI{1123}{\kelvin} under Ar atmosphere. After the \SI{48}{\hour} treatment, dark grey pellets were obtained which are reasonably stable in air.

Polycrystalline samples of 1111 compounds were obtained by solid state reaction of the stoichiometric mixture of REAs (RE: La, Sm), RESb, Fe\chem{2}O\chem{3}, REF\chem{3} and Fe powders. The mixture was prepared in an Ar filled glovebox, pressed into pellets, sealed in quartz tubes with an Ar pressure of \SI{0.05}{\bar} and heated in a box-type furnace with a rate of \SI{50}{\kelvin \per \hour} to \SI{1373}{\kelvin}. After \SI{48}{\hour} the furnace was switched off and cooled down to room temperature. After the heat treatment dark grey pellets were obtained, which are  reasonably stable in air. The rare-earth arsenide and antimonide precursors were prepared by heat treatment of a stoichiometric mixture of the elements at \SI{670}{\kelvin} for \SI{12}{\hour} in an evacuated quartz tube.

The samples were characterized at room temperature by X-ray powder diffraction using a Seifert XRD 3000 diffractometer in Bragg-Brentano geometry with CuK$_{\alpha}$ radiation and secondary Ge(002) monochromator. Data was collected from 10$^\circ$ - 100$^\circ$ in steps of 0.02$^\circ$. To determine the lattice parameters and the atomic positions, we used the LeBail method and Rietveld refinement, respectively. Crystal quality was tested by SEM/EDX using a Zeiss LEO 1530 (FE-SEM with Schottky-field-emission ca\-thode, in-lens detector, Back Scattered Electron Detector) using an accelerating voltage of \SI{10}{\kilo \volt}. The EDX data was collected by an INCA Energy System from Oxford Instruments. The temperature dependence of dc resistivity was measured with a standard four probe technique. Small rods with a length of \SI{5}{\milli \metre} and thickness of \SI{1}{\milli \metre} were cut from the pellets and four copper wires were connected with conductive silver. To confirm the occurrence of superconductivity, we used a home made magnetic ac-susceptometer and measured the screening signal at a frequency of \SI{2.7}{\kilo \hertz}.

\section{Results}

\subsection*{Ba122 samples}

\begin{figure}
\includegraphics[width=0.45\textwidth,keepaspectratio]{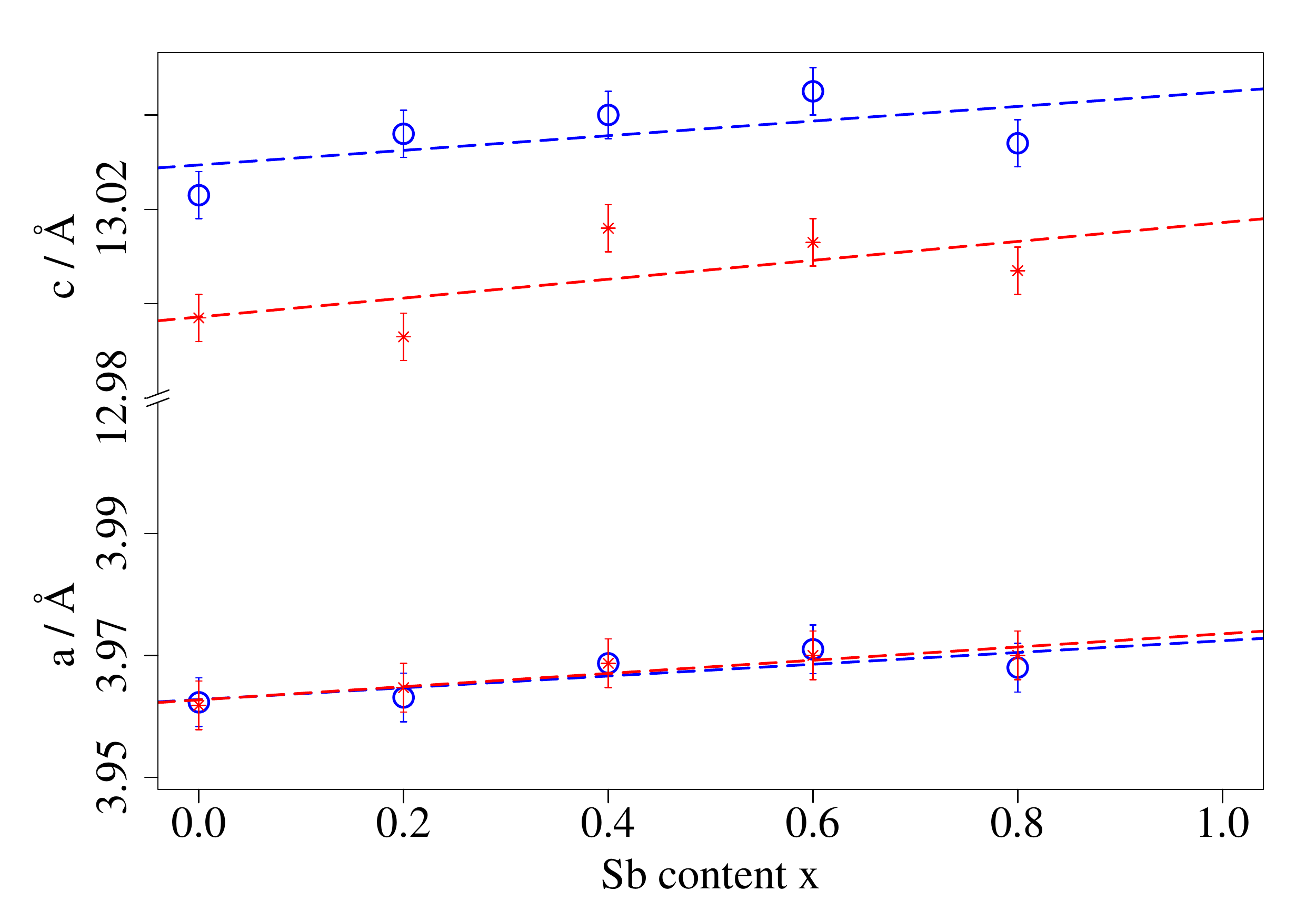}
\caption{Lattice parameters versus Sb content in BaFe\chemex{2}{y}Co\chem{y}As\chemex{2}{x}Sb\chem{x}. Blue: non-superconducting samples ($y$ = 0) with SDW-transition; Red: superconducting compounds ($y$ = 0.17). The tiny trend of increasing lattice parameters is within the standard deviation errors and therefore no proof for a successful Sb substitution in the 122-compounds.}
\label{fig:Ba122_latt}
\end{figure}

The powder patterns of non-superconducting cobalt-free samples with composition BaFe\chem{2}As\chemex{2}{x}Sb\chem{x} could be well indexed with the tetragonal ThCr\chem{2}Si\chem{2}-type structure of space-group I4/mmm. The lattice parameters of the Sb-free samples $a$ = \SI{3.9623(2)}{\angstrom} and $c$ = \SI{13.024(1)}{\angstrom}  agree with the results published by Rotter et al. \citep{rotter_spin-density-wave_2008} and, judged from powder X-ray results, the samples are single-phase. For the antimony free sample we obtain a residual from  Rietveld refinement of $R_\mathrm{p}$ = \SI{4.49}{\percent}. 
The lattice parameters and unit cell volume versus nominal Sb concentration of the substitution series are plotted in Fig. \ref{fig:Ba122_latt}. A slight increase of the lattice parameters with $x$ is suggested, however, this increase is smaller than the standard deviation of the measured values and thus cannot be taken as evidence for the solubility of antimony in this structure. In order to confirm this, we released the occupation numbers of arsenic and antimony in the refinement. For the $x = 0$ and $x = 0.4$ samples we found no occupation of antimony at the arsenic site. There is an increase of the residual to $R_\mathrm{p}$ = \SI{10.47}{\percent} at $x = 0.4$. In the qualitative phase analysis we found impurity phases up to an impurity level of \SI{20}{\percent}. We were able to identify iron arsenide and iron antimonide as well as barium silicate which appears to be a reaction product of the alcaline earth with the quartz glass ampoule.

The  temperature dependent measurement of resistivity down to \SI{4.2}{\kelvin} for the antimony free sample reveals a transition at \SI{136(5)}{\kelvin} related to the SDW transition. This transition temperature is not affected by increasing the nominal antimony concentration.

The powder patterns of the Co-substituted superconducting samples BaFe\chem{1.83}Co\chem{0.17}As\chemex{2}{x}Sb\chem{x} could also be indexed with the ThCr\chem{2}Si\chem{2}-type structure. The antimony free sample had also a single phase X-ray pattern. The lattice parameters of the antimony free sample are  $a = \SI{3.9618(2)}{\angstrom}$ and $c = \SI{12.997(1)}{\angstrom}$, slightly smaller than the parameters of the non-superconducting $x = 0$ sample, indicative of a successful cobalt substitution. The residual is $R_\mathrm{p} = \SI{6.77}{\percent}$. 
The lattice parameters of samples with higher nominal antimony content remain unchanged within standard deviation errors. However, impurity phases appear in the powder patterns and there is also an increase of the residual.

The temperature dependent electrical resistivity down to \SI{4.2}{\kelvin} shows a drop to zero at $T_\mathrm{c} = \SI{24}{\kelvin}$. The appearance of superconductivity was confirmed by measuring the screening signal in ac-susceptibility, which confirmed the transition temperature, in good agreement with literature \citep{rotter_spin-density-wave_2008, nakajima_possible_2009}. With higher nominal antimony concentrations we found no change in transition temperatures neither in the Co-free SDW samples, nor in the Co-substituted superconducting samples.

\subsection*{La1111 samples}

\begin{figure}[tb]
\includegraphics[width=0.45\textwidth,keepaspectratio]{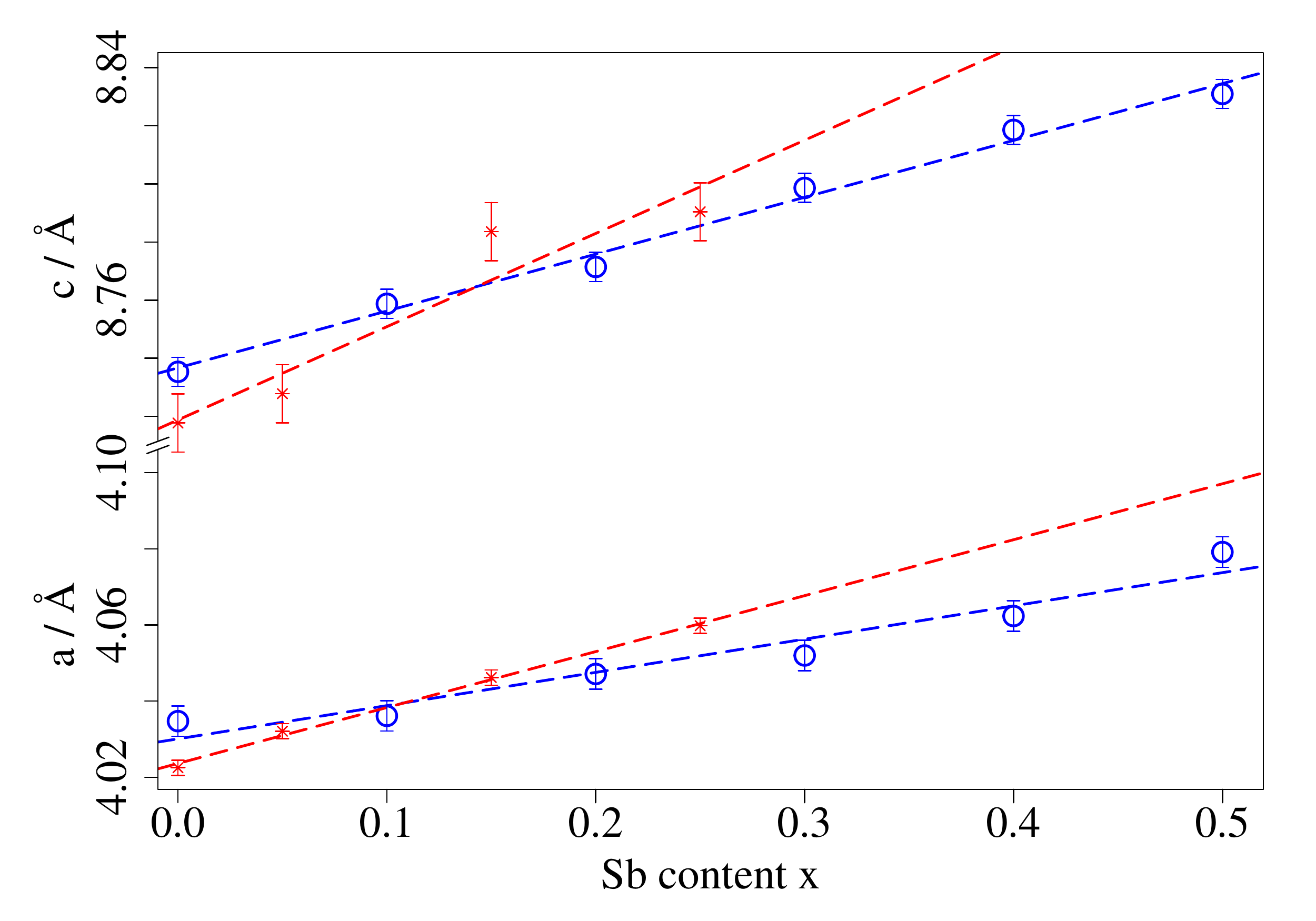}
\caption{Lattice parameters of LaFeAs\chemex{1}{x}Sb\chem{x}O\chemex{1}{y}F\chem{y} versus the nominal antimony content. Blue: Non-superconducting ($y$ = 0) compounds; Red: superconducting ($y$ = 0.15) compounds. The parameters follow Vegard's law (see eqns. 1--4). For the non-superconducting compounds the limit of solubility lies at $x = 0.5$ and for the superconducting compounds at $x = 0.25$.}
\label{fig:La1111_latt}
\end{figure}

In figure \ref{fig:La1111_latt} the variation of the lattice parameters with  antimony content is shown. Shown in blue are the results for the LaFeAs\chemex{1}{x}Sb\chem{x}O series with SDW character and in red  for the superconducting LaFeAs\chemex{1}{x}Sb\chem{x}O\chem{0.85}F\chem{0.15} samples. At room temperature, the diffraction patterns could be well indexed by a tetragonal ZrCuSiAs structure with space group P4/nmm. 

The $R_\mathrm{p}$ factor for the non-superconducting antimony free sample was determined to \SI{4.66}{\percent}, indicative for a single phase sample. The lattice parameters were $a = \SI{4.0348(1)}{\angstrom}$ and $c = \SI{8.7362(4)}{\angstrom}$. The lattice parameters increase with increasing antimony content following Vegard's law between an antimony content of 0 and 0.5 with:
\begin{eqnarray}
a &=& \left[0.087(11) \cdot x + 4.040(3)\right] \si{\angstrom}\\
c &=& \left[0.196(10) \cdot x + 8.737(3)\right] \si{\angstrom}
\end{eqnarray}
For higher nominal antimony concentrations $x>0.5$, no  homogeneous samples were obtained. The lattice parameters of the ZrCuSiAs-type phase do not increase further, instead, impurity phases appear.

The $R_\mathrm{p}$ factor of the antimony free superconducting sample is \SI{6.39}{\percent}, the lattice parameters are $a = \SI{4.0225(1)}{\angstrom}$ and $c = \SI{8.7177(4)}{\angstrom}$. Between an antimony content of 0 and 0.25 the lattice parameters follow Vegard's law, well described by: 
\begin{eqnarray}
a &=& \left[0.146(7) \cdot x + 4.023(1)\right]\si{\angstrom}\\
c &=& \left[0.32(7) \cdot x + 8.72(1)\right]\si{\angstrom}
\end{eqnarray}
For higher nominal antimony concentrations $x>0.25$, no homogeneous samples were obtained.

\begin{figure}[tb]
\includegraphics[width=0.45\textwidth,keepaspectratio]{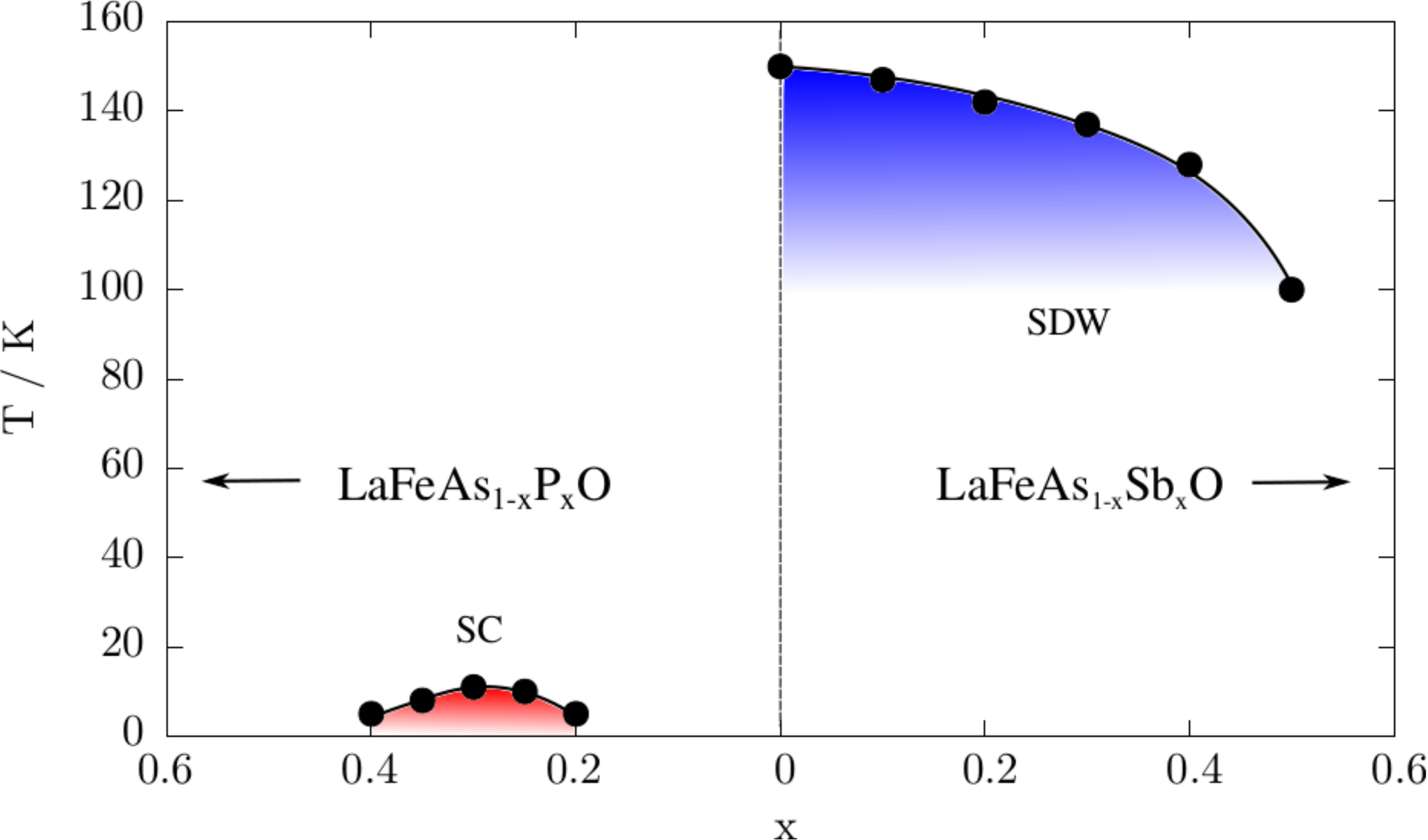}
\caption{Right: Mean value of the crystallographic/SDW transition (see text) versus  antimony content $x$. The part shaded in blue shows the temperatures measured in the present work. No superconductivity was found down to 4.2\,K. Left: The red part shows the superconducting regime of phosphorus doped compounds, taken from \citep{wang_superconductivity_2009}.}
\label{fig:phase_diagr_1111}
\end{figure}

The non-superconducting samples show two transitions in the resistivity signal similar to those described in \citep{klauss_commensurate_2008}. The first transition refers to the structural phase transition from a tetragonal to an orthorhombic structure. For the $x = 0$ sample we found the transition temperature $T_\mathrm{S} = \SI{160}{\kelvin}$. The second transition refers to the N\'{e}el temperature $T_\mathrm{N} = \SI{135}{\kelvin}$. The temperature difference between the structural and magnetic transition is constant over the investigated antimony range, thus, for the sake of simplicity, we plot the mean of the transition temperatures $(T_\mathrm{S}+T_\mathrm{N})/2$ in figure \ref{fig:phase_diagr_1111}, labelled SDW and emphasized by  blue color. With higher antimony concentration, the transition temperatures decrease. Within the homogeneous range we found a minimum mean transition temperature of \SI{100}{\kelvin} for $x = 0.5$ but no superconductivity down to \SI{4.2}{\kelvin}. For higher Sb concentration, the semiconducting behaviour of impurities dominates the resistivity signal and it became impossible to extract transition temperatures for the LaFeAsO phase.

\begin{figure}[tb]
\includegraphics[width=0.45\textwidth,keepaspectratio]{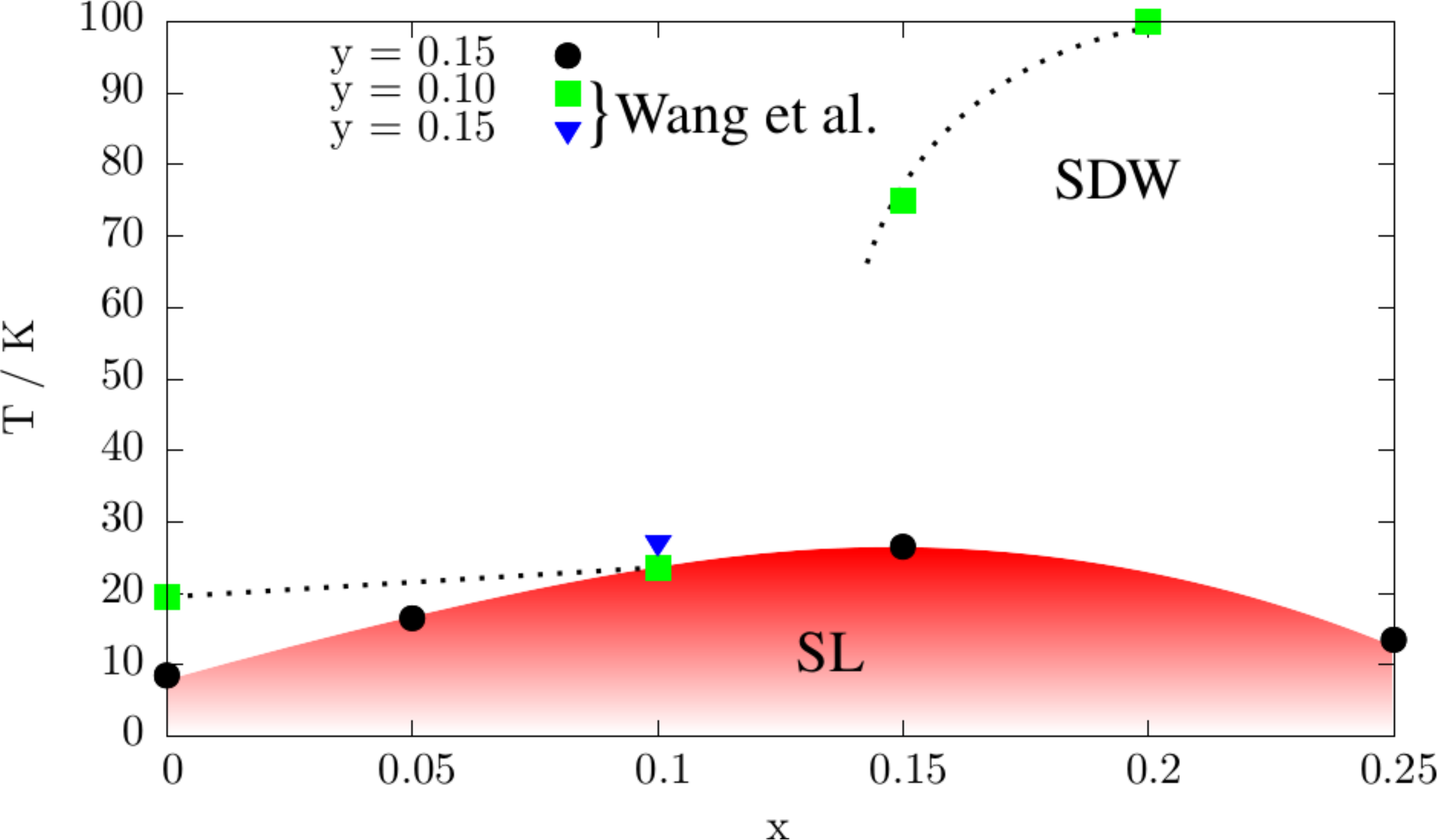}
\caption{Mean superconducting transition temperatures of LaFeAs\chemex{1}{x}Sb\chem{x}O\chemex{1}{y}F\chem{y} with $y=0.15$ versus antimony content $x$ (black points). We find a maximum $T_\mathrm{c}$ of \SI{26.5}{\kelvin} in this system and no SDW transition.  Squares (plotted in green) and triangles (in blue) are taken from Wang et al. \citep{wang_structural_2010}}
\label{fig:phase_diagr_1111of}
\end{figure}

With a fluorine content of $y = 0.15$, the samples are superconducting. For Sb-free samples we found a transition temperature of $T_\mathrm{c} = \SI{9}{\kelvin}$ which agrees with the phase diagram suggested in Oka et al. \citep{oka_antiferromagnetic_2012}. With increasing antimony content, the transition temperature increases up to a maximum $T_\mathrm{c}$ of \SI{28}{\kelvin} in the resistivity signal and \SI{25}{\kelvin} in the ac susceptibility signal. The black data points in figure \ref{fig:phase_diagr_1111of} show the mean transition temperature of our resistivity and susceptibility measurements. The additional points plotted in green and in blue are taken from Wang et al. \citep{wang_structural_2010}.

\subsection*{Sm1111 samples}

When lanthanum is replaced with samarium, the cell volume of the 1111 phase is smaller. We measured $a = \SI{3.939(5)}{\angstrom}$ and $c = \SI{8.49(1)}{\angstrom}$  in good agreement with \citep{chen_superconductivity_2008}. However, with increasing nominal antimony content, merely the impurity level increases and the crystallographic and physical properties of the 1111 phase remain constant. Thus, we conclude that there is no solubility of antimony in  SmFeAs\chemex{1}{x}Sb\chem{x}O.

\section{Discussion}

The invariance of the crystallographic and physical properties in the 122-compounds against nominal Sb concentration as well as direct structure refinements prove the absence of solubility of Sb in this structure. BaFe$_2$As$_2$ has the highest unit cell volume in the class of known 122 iron arsenides. It is thus unlikely that Sb can be substituted into 122 iron pnictide compounds with smaller unit cell volume. The unit cell size of BaFe$_2$As$_2$ appears to be at the stability limit of this class of compounds.

The measured lattice parameters of the non-superconducting La1111 parent compounds agree well with the values computed in \citep{lebegue_delicate_2009} or \citep{moon_enhanced_2008}. However, the slope of the linear regression with Vegard's law differs from the measured data. In \citep{lebegue_delicate_2009} the slope of $a$ is \SI{0.1271}{\angstrom} and \SI{0.6078}{\angstrom} for $c$ and in \citep{moon_enhanced_2008} it is \SI{0.107}{\angstrom} for $a$ and \SI{0.551}{\angstrom} for $c$. Our smaller slope of \SI{0.087}{\angstrom} for $a$ and \SI{0.196}{\angstrom} for $c$ may be a result of increasing disorder which was neglected in the calculation of the lattice parameters in the theoretical studies.
We find that La1111 compounds show a solubility up to an antimony concentration of $x = 0.5$ in the non-superconducting compounds confirming \citep{carlsson_effect_2011}
and $x = 0.25$ in the superconducting fluorine doped compounds with $y = 0.15$. Carlsson et al. found slightly larger lattice parameter for the non-superconducting Sb-substituted compounds but the same trend in the the SDW transition temperature \citep{carlsson_effect_2011}. With increasing lattice parameters, that is negative chemical pressure, the interlayer distance between the FeAs layers increases due to the larger $c$ axis and thus the magnetic interaction is weakened. 
As a consequence, the SDW transition temperature decreases from \SI{150}{\kelvin} for $x=0$ to \SI{100}{\kelvin} for $x=0.5$. 

With higher antimony concentration, no single phase samples were obtained. We conclude that there is a solubility limit for antimony near $x = 0.5$. In contrast to the phosphorus substituted compounds \citep{wang_superconductivity_2009}, we found no superconductivity in the antimony substituted compounds, in agreement with \citep{carlsson_effect_2011}. The SDW cannot be suppressed and superconductivity cannot emerge because of the enhanced Hund's rule coupling discussed in \citep{moon_enhanced_2008}. The Fe $d$ states are localized since the orbital overlap is reduced due to the larger iron-pnicogen distance.

The results on the fluorine doped samples reveal an increase of the superconducting transition temperature with increasing Sb concentration. A problem is  the exact determination of fluorine content. We have compared the transition temperature of our $x = 0$ compound with the phase diagram from \citep{oka_antiferromagnetic_2012} and find agreement with the 0.15 fluorine doping level. Another benchmark is the point for $x = 0.1, y=0.15$ from \citep{wang_structural_2010} which agrees with our suggested phase diagram (blue triangle in Fig. \ref{fig:phase_diagr_1111of}). With our compounds and measurement techniques, we found a maximum $T_\mathrm{c}$ of \SI{27}{\kelvin} in the LaFeAs\chemex{1}{x}Sb\chem{x}O\chem{0.85}F\chem{0.15} system. In contrast to the work of Wang et al. (green squares in Fig. \ref{fig:phase_diagr_1111of}), we found no recurrence of the SDW transition with higher antimony content. For concentrations higher than $x = 0.25$ we obtain no single phase compounds.

The smaller unit cell volume of the samarium compounds leads to a blockade of solubility of antimony. We found no variation of crystallographic and physical properties with nominal antimony content. Obviously, there is no solubility of antimony in the Sm1111-system.

\section{Conclusion}

We have investigated the solubility of Sb in some iron arsenides of the 122 and 1111 type. In the case of fluorine-free  LaFeAs$_{1-x}$Sb$_x$O compounds, the substitution was possible up to a solubility limit $x=0.5$. The   SDW transition temperature decreases with increasing Sb concentration. The solubility limit of Sb in the superconducting fluorine-doped compounds LaFeAs$_{1-x}$Sb$_x$O$_{0.85}$F$_{0.15}$ 
is $x=0.25$. We found a maximum $T_\mathrm{c}$ of \SI{27}{\kelvin} at $x=0.15$. 
No antimony substitution of arsenic was possible in the Sm1111 and the Ba122 compounds.  With the application of ``negative chemical pressure'', these compounds become unstable against the formation of foreign phases.

\section*{Acknowledgements}
We thank M. Heider from the "Bayreuther Institut f\"ur Ma\-kro\-mo\-lek\"ul\-for\-schung" for the SEM/EDX measurements. N. Kurz, L. Wehmeier and S. Wolf for their help with sample preparation and last but not least C. Kerling for the technical support in our lab. 

\bibliographystyle{elsarticle-num}

\bibliography{bibliography}

\end{document}